\documentclass[twocolumn,preprintnumbers,superscriptaddress,unsortedaddress]{revtex4}
\usepackage{amssymb}
\usepackage{amsmath}
\usepackage{graphicx}
\usepackage{dcolumn}
\usepackage{bm}
\usepackage{amsfonts}
\usepackage{units}
\usepackage{hyperref}

\UseRawInputEncoding

\begin{document}

\title{Optical interface for a hybrid magnon-photon resonator}
\author{Banoj Kumar Nayak*\footnotetext{*These authors contributed equally to this work.}}
\affiliation{Andrew and Erna Viterbi Department of Electrical Engineering, Technion,
	Haifa 32000 Israel}
\author{Cijy Mathai*}
\affiliation{Andrew and Erna Viterbi Department of Electrical Engineering, Technion,
	Haifa 32000 Israel}
\author{Dekel Meirom}
\affiliation{Andrew and Erna Viterbi Department of Electrical Engineering, Technion,
	Haifa 32000 Israel}
\author{Oleg Shtempluck}
\affiliation{Andrew and Erna Viterbi Department of Electrical Engineering, Technion,
	Haifa 32000 Israel}
\author{Eyal Buks}
\affiliation{Andrew and Erna Viterbi Department of Electrical Engineering, Technion,
	Haifa 32000 Israel}
\date{\today }

\begin{abstract}
We study optical detection of magnetic resonance of a ferrimagnetic sphere
resonator, which is strongly coupled to a microwave loop gap resonator.
Optical fibers are employed for coupling the sphere resonator with light in
the telecom band. We find that magnetic resonance can be optically detected
in the region of anti-crossing between the loop gap and the ferrimagnetic
resonances. By measuring the response time of the optical detection we rule
out the possibility that microwave induced heating is responsible for the
optical detectability.
\end{abstract}

\maketitle

Magnons are widely employed in a variety of devices \cite%
{Hill_S227,Lecraw_1311,
Kumar_435802,Sander_363001,Vedmedenko_453001,Chumak_244001,rezende2020fundamentals}%
, including narrow band oscillators \cite{Ryte_434}, filters \cite{Tsai_3568}%
, and parametric amplifiers \cite{Kotzebue_773}. Magnons can couple with
microwave (MW) photons \cite{Zhang_156401,Zhu_2005_06429}, optical photons 
\cite{Bai_1,Demokritov_441,Zhang_123605, Osada_223601,
Stancil_Spin,Kajiwara_262,Haigh_133602,Sharma_094412,Hisatomi_207401,Pantazopoulos_104425,Hisatomi_174427}%
, phonons \cite{Sinha_432,Zhang_e1501286}, and with superconducting qubits 
\cite{Tabuchi_405,Lachance_e1603150,Lachance_070101,Wolski_2005_09250}.
Hybrid magnon devices may help developing optical channels linking remote
quantum computers \cite{Kaur_032404,Huebl_127003,Tabuchi_083603}.

Here we study a hybrid system composed of a MW loop gap resonator (LGR)
strongly coupled to a ferrimagnetic sphere resonator (FSR) made of yttrium
iron garnet (YIG) \cite{Cherepanov_81,Serga_264002}. Optical fibers are
employed for transmitting light in the telecom band through the sphere. The
frequency of the hybrid FSR-LGR system is controlled using an externally
applied magnetic field (generated by a magnetized Neodymium). We explore
magneto-optic (MO) coupling and Faraday rotation of optical polarization,
and demonstrate optical detection of magnetic resonance (ODMR) of the hybrid
FSR-LGR system. ODMR of FSR has been demonstrated before in \cite%
{Chai_021101}, by coupling a tapered optical fiber to whispering gallery
modes of an FSR. However, the ODMR method that has developed in \cite%
{Chai_021101} is based on heating induced by MW driving, and consequently
the response time of this method is relatively long (on the order of a
second). As is shown below, the response time of our ODMR method, which is
not based on heating, is significantly shorter (limited by the ring down
time of the FSR, which is about $1\unit{%
\mu%
s}$).

\begin{figure}[tbp]
\begin{center}
\includegraphics[width=3.2in,keepaspectratio]{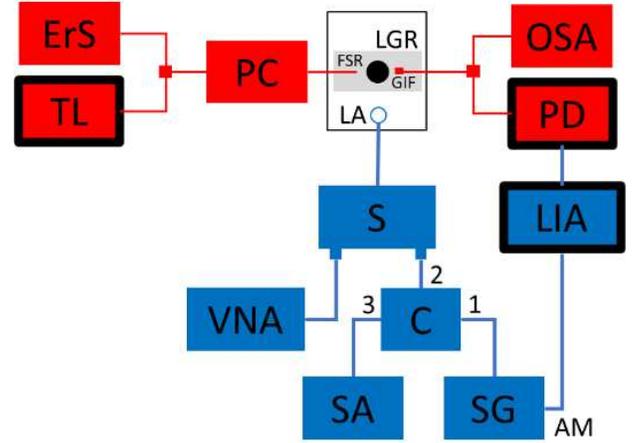}
\end{center}
\caption{{}Experimental setup. Optical components [ErS (Erbium source), TL
(tunable laser), PC (polarization controller), OSA (optical spectrum
analyzer) and PD (photodetector)] and fibers are red colored, and MW
components [LA (loop antenna), S (splitter), C (circulator), VNA (vector
network analyzer), SA (spectrum analyzer), SG (signal generator), and LIA
(lockin amplifier)] and coaxial cables are blue colored. The LA is weakly
coupled to the FSR-LGR system. Optical fibers are installed on both sides of
the FSR for transmission of light through the sphere. Components outlines by
a thick black line (TL, PD and LIA) are used only for the measurements
presented in Fig. \protect\ref{FigTL}.}
\label{FigSetup}
\end{figure}

The experimental setup, which is schematically shown in Fig. \ref{FigSetup},
is designed to allow exploring the MO coupling between MW and optical
photons, which is mediated by FSR magnons. In Fig. \ref{FigSetup}, optical
components and fibers are red colored, whereas blue color is used to label
MW components and coaxial cables.

A MW cavity made of an LGR allows achieving a relatively large coupling
between magnons and MW photons \cite{Froncisz_515,Zhang_205003,Mathai_054428}%
. The LGR is fabricated from a hollow concentric aluminium tube. A sapphire
strip of $260\unit{%
\mu%
m}$ thickness is inserted into the gap in order to increase its capacitance,
which in turn reduces the frequency $f_{\mathrm{c}}$ of the LGR fundamental
mode \cite{krupka_387}. An FSR made of YIG having radius of $R_{\mathrm{s}%
}=125\unit{%
\mu%
m}$ is held by two ceramic ferrules inside the LGR. The applied static
magnetic field $\mathbf{H}$ is controlled by adjusting the relative position
of the magnetized Neodymium using a motorized stage. The LGR-FSR coupled
system is encapsulated inside a metallic rectangular shield made of aluminum
(represented by the black colored rectangle in Fig. \ref{FigSetup}). The
cavity is weakly coupled to a loop antenna (LA). More information about the
FSR-LGR hybrid system, including its fabrication and magnetic energy density
distribution, can be found in Ref. \cite{Mathai_054428}.

\begin{figure*}[tbp]
\begin{center}
\includegraphics[width=6.4in,keepaspectratio]{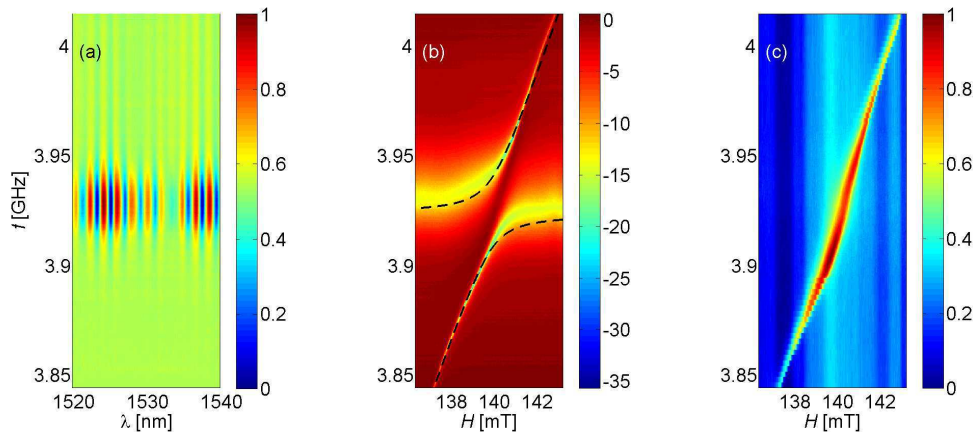}
\end{center}
\caption{Continuous wave measurements. In both (a) and (c) the ErS is
employed as a source, and the optical transmission measurement is performed using the OSA. (a)
Measured optical spectrum data as a function of SG frequency $f$ with SG power of $20$~dBm relative to measured optical spectrum corresponding to smallest SG frequency in the plot, with a fixed magnetic field of $H=140.13 \unit{mT}$%
, showing Fabry--P\'{e}rot oscillation near the FSR-LGR resonance. (b) VNA
reflection $\left\vert S_{11}\right\vert ^{2}$ in dB units as a function of
frequency $f$ and magnetic field $H$ with input power of $-30$~dBm. (c)
Optical intensity (in arbitrary units) measured at a specific
wavelength of $\protect\lambda =1524.292\unit{nm}$ as a function of SG
frequency $f$ and magnetic field $H$, with SG power of $20$~dBm.}
\label{FigEr}
\end{figure*}

A vector network analyzer (VNA) is employed for measuring the MW reflection
coefficient $\left\vert S_{11}\right\vert ^{2}$. The plot shown in Fig. \ref%
{FigEr}(b) exhibits $\left\vert S_{11}\right\vert ^{2}$ in dB units as a
function of the externally applied magnetic field $H$ and VNA frequency $f$.
The measurement is performed in the region of anti-crossing between the LGR
fundamental mode at frequency $f_{\mathrm{c}}=3.9235\unit{GHz}$ and the
Kittel mode FSR frequency $f_{\mathrm{s}}$ , which is given by $f_{\mathrm{s}%
}=\gamma _{\mathrm{g}}H/\left( 2\pi \right) $, where $\gamma _{\mathrm{g}%
}/2\pi =27.98~\unit{GHz}~\unit{T}^{-1}$\ is the gyromagnetic ratio \cite%
{Jin_thesis,Fletcher_687}.

The frequencies $f_{\pm }$ of the hybrid FSR-LGR eigen modes are given by 
\cite{Sainz_de_los_Terreros_1906}%
\begin{equation}
f_{\pm }=\frac{f_{\mathrm{c}}+f_{\mathrm{s}}}{2}\pm \sqrt{\left( \frac{f_{%
\mathrm{c}}-f_{\mathrm{s}}}{2}\right) ^{2}+g^{2}}\;.  \label{omega pm}
\end{equation}%
where $g$ is the FSR-LGR coupling coefficient \cite{Wang_224410,Mathai_67001}%
. The overlaid black dashed lines in Fig. \ref{FigEr}(b) are calculated
using Eq. (\ref{omega pm}). A fitting procedure yields the value $g/\left(
2\pi \right) =16\unit{MHz}$. Note that in general, $g$ is proportional to
the FSR volume.

In the telecom band (wavelength $\lambda \simeq 1.5\unit{%
\mu%
m}$) YIG has an optical absorption coefficient $\alpha $ of about $\alpha
=\left( 0.5\unit{m}\right) ^{-1}$ \cite{Wood_1038}, a Verdet constant $A_{%
\mathrm{V}}$ of about $A_{\mathrm{V}}=5\times 10^{-5}\unit{mT}^{-1}\unit{%
\mu%
m}^{-1}$ \cite{Onbasli_1,Jooss_651}, and a polarization beating length $l_{%
\mathrm{P}}$ in magnetic saturation of about $l_{\mathrm{P}}=7.\,0\unit{mm}$ 
\cite{Zhang_591,Onbasli_1,Donati_372}. YIG spheres can be employed for
making optical circulators and isolator in the telecom band \cite%
{Yokohama_746l,Okamoto_36}. An optical cavity can be constructed to enhance
Faraday rotation \cite{Stone_849,Chang_8526}.

In our setup telecom light is transmitted through the FSR using single mode
optical fibers. The FSR serves as a thick optical lens having focal length
of $F_{\mathrm{FSR}}=\left( 1/2\right) n_{0}R_{\mathrm{s}}/\left(
n_{0}-1\right) $, where $n_{0}=2.19$ is YIG refractive index in the telecom
band \cite{Okamoto_36,Yokohama_746}, A graded index fiber (GIF) is attached
to the tip of one the the fibers that are installed near the FSR (see Fig. %
\ref{FigSetup}). The length of the GIF is $0.1p_{\mathrm{g}}$, where $p_{%
\mathrm{g}}=1\unit{mm}$ is the GIF pitch. Focusing is achieved by displacing
\ both fibers along the optical axis and maximizing the fiber to fiber
transmission $T_{\mathrm{F}}$, which for the current device under study is $%
T_{\mathrm{F}}=0.59$.

Spontaneous emission from an Erbium doped fiber is used as the optical
source for the measurements presented in Fig. \ref{FigEr}. The Erbium source
(ErS) intensity peaks near wavelength of $1530\unit{nm}$. A polarization
controller (PC) is employed to manipulate the light transmitted through the
FSR. An optical spectrum analyzer (OSA) having resolution of $0.004\unit{nm}$
is used to probe the transmitted light. All fibers are single mode having $%
125\unit{%
\mu%
m}$ clad diameter and $9\unit{%
\mu%
m}$ core diameter.

\begin{figure*}[tbp]
\begin{center}
\includegraphics[width=6.4in,keepaspectratio]{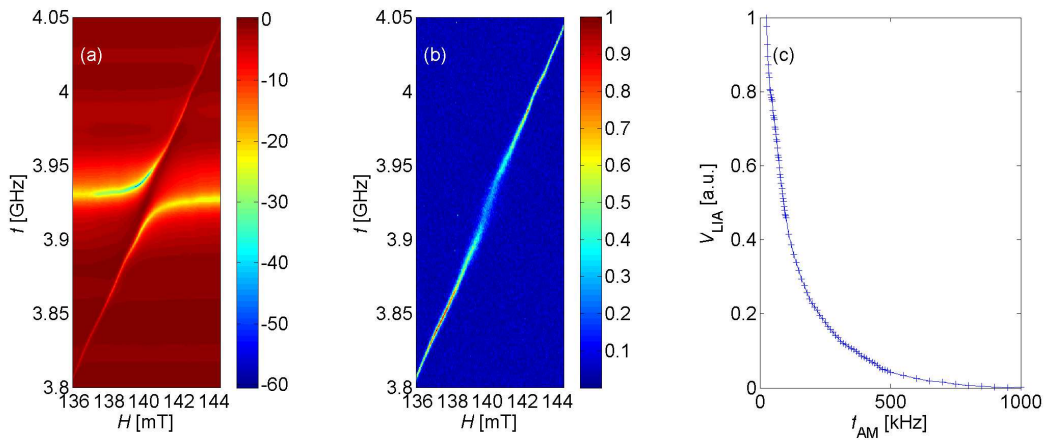}
\end{center}
\caption{LIA measurements. (a) VNA reflection $\left\vert S_{11}\right\vert
^{2}$ in dB units as a function of frequency $f$ and magnetic field $H$. The
VNA input power is $-10$~dBm. (b) LIA measured voltage amplitude (in
arbitrary units) as a function of SG frequency $f$ and magnetic field $H$,
with SG power of $-10$~dBm. In both (a) and (b) the TL wavelength is $1530.87%
\unit{nm}$ and power is $-2.6$~dBm. (c) The dependence of LIA measured
voltage amplitude $V_{\mathrm{LIA}}$ (in arbitrary units) on LIA modulation
frequency $f_{\mathrm{AM}}$. For this measurement, the TL wavelength is set to $1538.556\unit{nm}$ and TL power is set to $6$~dBm.}
\label{FigTL}
\end{figure*}

The OSA is employed for probing the transmitted light in the range $1520%
\unit{nm}$ to $1540\unit{nm}$, as a function of MW driving frequency $f$
applied to the LA, with a fixed magnetic field of $140.13\unit{mT}$ [see Fig. \ref%
{FigEr}(a)]. In this measurement, a signal generator (SG) operating at $20$~dBm serves as a source [see Fig. \ref{FigSetup}, and note that a circulator
(C) and a MW spectrum analyzer (SA) are used to probe the back reflected MW
signal]. The measured optical transmission shown in Fig. \ref{FigEr}(a)
reveals Fabry--P\'{e}rot oscillation near the FSR resonance $f_{\mathrm{s}}$%
. The wavelength period of the Fabry--P\'{e}rot oscillation is observed to
be $1.77\unit{nm}$. The oscillation is attributed to an optical cavity formed
between both fibers coupled to the FSR due to Fresnel reflection at the
fibers' tips. The measured spacing of $1.77\unit{nm}$ allows estimating the
distance between the fibers to be $700\unit{%
\mu%
m}$.

To study the dependence on both the MW frequency $f$ as well as magnetic
field $H$, the optical intensity is recorded at wavelength $1524.292\unit{nm}
$ [see Fig. \ref{FigEr}(c)], at which the transmission is maximized [see
Fig. \ref{FigEr}(a)]. The SG frequency is varied from $3.8\unit{GHz}$ to $%
4.05\unit{GHz}$, and the power is set to $20$~dBm. The measured optical
intensity peaks near the Larmor resonance, i.e. when $f\simeq f_{\mathrm{s}}$%
. Note that the splitting between $f_{+}$ and $f_{-}$ cannot be resolved in
Fig. \ref{FigEr}(c) due to anisotropy-induced Kerr nonlinearity \cite%
{Mathai_054428}.

Next we explore the response time of the above-discussed ODMR method. This
is done in order to determine the role played by MW induced heating, which
has a relatively long time scale \cite{Chai_021101}. To that end, we perform
experiments using a lockin amplifier (LIA). Components outlines by a thick
black line in the setup sketch shown in Fig. \ref{FigSetup} (TL, PD and LIA)
are used only for the LIA measurements presented in Fig. \ref{FigTL}. A
tunable laser (TL) is used instead of the high bandwidth ErS. The OSA is
replaced with a photodetector (PD) to measure the optical intensity. The LIA
reference signal is used to amplitude modulate (AM) the SG signal at a
modulation frequency $f_{\mathrm{AM}}$, and the PD signal output is fed into
the LIA input port. For LIA measurement shown in Fig. \ref{FigTL}(c), the tunable laser wavelength is set to $1538.556\unit{nm}
$,  which corresponds to the second highest optical intensity in the
spectrum shown in Fig. \ref{FigEr}(a).

Both the VNA measurement shown in Fig. \ref{FigTL}(a) and the LIA
measurement shown in Fig. \ref{FigTL}(b) are performed with MW power of $-10$~dBm and TL optical power of $-2.6$~dBm. Figure \ref{FigTL}(c) shows a plot
of LIA voltage amplitude $V_{\mathrm{LIA}}$ (in arbitrary units) as a
function of modulation frequency $f_{\mathrm{AM}}$. The measured dependency
on $f_{\mathrm{AM}}$ indicates that the ODMR response time is on the order
of a microsecond. This observation suggests that the response time is limited by FSR
damping, and it rules out the possibility that heating plays a dominant role
in the underlying mechanism allowing the ODMR.

In summary, ODMR of FSR is demonstrated in the telecom band, and the
possibility that MW induced heating is the underlying mechanism is ruled
out. The ODMR method is compatible with ultra low temperatures (due to the
very low optical absorption of YIG in the telecom band), and thus it may
help developing an optical interface for superconducting qubits \cite%
{Tabuchi_405,Lachance_e1603150,Lachance_070101,Wolski_2005_09250}.

This work was supported by the Israeli science foundation, the Israeli
ministry of science, and by the Technion security research foundation.

The data that support the findings of this study are available from the corresponding author upon reasonable request.

\bibliographystyle{IEEEtran}
\bibliography{Eyal_Bib}

\end{document}